\documentclass[a4paper,11pt]{article}
\usepackage{jheppub} 
\usepackage{epsfig}
\usepackage{graphicx}
\usepackage[T1]{fontenc}
\def\be{\begin{equation}}
\def\ee{\end{equation}}
\def\bea{\begin{array}}
\def\eea{\end{array}}
\def\beqa{\begin{eqnarray}}
\def\eeqa{\end{eqnarray}}
\def\beqas{\begin{eqnarray*}}
\def\eeqas{\end{eqnarray*}}

\def\bp{\begin{picture}}
\def\ep{\end{picture}}
\def\bc{\begin{center}}
\def\ec{\end{center}}
\def\bfig{\begin{figure}}
\def\efig{\end{figure}}

\def\bit{\begin{itemize}}
\def\eit{\end{itemize}}
\def\nn{\nonumber}
\def\f{\frac}

\def\[{\left[}
\def\]{\right]}
\def\({\left(}
\def\){\right)}

\def\..{\left.}
\def\.{\right.}
\def\tl{\tilde}
\def\ra{\rightarrow}

\def\tm{\times}

\def\al{\alpha}

\def\ep{\epsilon}

\def\Ga{\Gamma}

\def\pa{\partial}
\def\pr{\prime}

\def\kev{\rm keV}
\title{\boldmath  Explaining The XENON1T Excess With Light Goldstini Dark Matter}
\author[a]{Junjie Cao,}
\affiliation[a]{Department of Physics, Henan Normal University, 453007, Xinxiang, P.R. China}
\author[b]{Xiao Kang Du,}\author[b]{Zhuang Li,}\author[b] {Fei Wang,}
\affiliation[b]{Department of Physics and Microelectronics, Zhengzhou University, 450000, ZhengZhou, P.R.China}
\author[c]{Yang Zhang}
\affiliation[c]{School of Physics and Astronomy, Monash University, Melbourne, Victoria 3800, Australia}
\emailAdd{feiwang@zzu.edu.cn}
\abstract{ In the scenario with a multiplicity of sectors which independently break supersymmetry, multiplicity of goldstini are predicted. We propose a new interpretation of the electron recoil excess at 2-7 keV observed in the XENON1T experiment with very long-lived goldstini DM elastically scattering off the electrons. The goldstini DM can be boosted by the late-decay of the other nearly degenerate (long-lived) goldstini DM, with their tiny mass difference being converted into kinetic energy of the lighter goldstini DM and neutrinos. We show that viable parameter space can be found which can explain the excess of electron recoil events around 2-3 keV recently reported by the XENON1T experiment.}
\begin{document}
\maketitle
\flushbottom

\section{Introduction}
Recently, the XENON1T experiment had observed an excess at 2-3 keV in their low energy electron recoil data with an exposure of 0.65 ton-years~\cite{Aprile:2020tmw}. There are 285 observed events over an expected background of $232\pm 15$ events within $1-7$ keV. Although the beta decay of background tritium can possibly contribute to such an excess, other explanations are still necessary because the  the background tritium content is not well understood. To interpret such an excess other than possible tritium sources,  three explanations are given in the report, namely the solar axion, the neutrino magnetic momentum and light bosonic dark matter(DM), respectively~\cite{Aprile:2020tmw}.  However, the preferred couplings for solar axion and neutrino magnetic momentum have already been ruled out by constraints from astrophysics. Regarding the potential importance of such an excess, it is instructive to seek other explanations.

Many alternative ideas had been proposed to explain the XENON1T excess, including the non-standard neutrino-electron interactions with light mediators~\cite{Boehm:2020ltd,Bally:2020yid,AristizabalSierra:2020edu,Khan:2020vaf,Jho:2020sku,Lindner:2020kko,Gao:2020wfr,
Ge:2020jfn,Coloma:2020voz,Miranda:2020kwy,Babu:2020ivd,Shoemaker:2020kji,Arcadi:2020zni}, absorption of axion or dark photon theories~\cite{Takahashi:2020bpq,Alonso-Alvarez:2020cdv,Choi:2020udy,DiLuzio:2020jjp,Buch:2020mrg,Nakayama:2020ikz,
An:2020bxd,Bloch:2020uzh,Budnik:2020nwz,Zu:2020idx,Gao:2020wer,DeRocco:2020xdt,Dent:2020jhf,Cacciapaglia:2020kbf,Sun:2020iim,Li:2020naa,Okada:2020evk,
Davighi:2020vap,Choi:2020kch,Long:2020uyf,Athron:2020maw,Chiang:2020hgb,Arias-Aragon:2020qtn}, the scattering of dark matter
candidates with electron~\cite{Kannike:2020agf,Fornal:2020npv,Su:2020zny,Chen:2020gcl,Bell:2020bes,Paz:2020pbc,Cao:2020bwd,Primulando:2020rdk,
Lee:2020wmh,Bramante:2020zos,Chao:2020yro,Ko:2020gdg,An:2020tcg,Alhazmi:2020fju,Baek:2020owl,Chigusa:2020bgq,He:2020wjs,Davoudiasl:2020ypv,
Choudhury:2020xui}, and the other mechansims~\cite{Harigaya:2020ckz,Dremin:2020dre,Dey:2020sai,Robinson:2020gfu,Dessert:2020vxy,
Bhattacherjee:2020qmv,Zioutas:2020cul,Croon:2020ehi}. We emphasize that explanations from cosmic dark matter with their
new features may shed new light on possible new physics beyond the Standard Model(SM), including the low energy supersymmetry(SUSY).

TeV scale SUSY is one of the most promising candidates for new physics beyond the SM. It can not only prevent the Higgs boson mass from acquiring dangerous quadratic divergence corrections, but also realize successful gauge coupling unification and provide viable DM candidates. The low energy SUSY spectrum is totally determined by the SUSY breaking mechanism, which can predict the low energy parameters by very few UV inputs. Depending on the way the SUSY breaking effects in the hidden sector communicate to the visible sector, the SUSY breaking mechanisms can be classified into gravity mediation~\cite{Chamseddine:1982jx,Nilles:1982ik,Ibanez:1982ee,Barbieri:1982eh,Nilles:1982dy,Ellis:1982wr,
Ellis:1983bp,Ohta:1982wn,Hall:1983iz,Wang:2018vrr,Wang:2015mea,Wang:2018vxp}, ~gauge mediation~\cite{Dine:1981za,Dimopoulos:1981au,Dine:1981gu,Dine:1993yw,Dine:1994vc,Dine:1995ag,
Giudice:1998bp}, anomaly mediation~\cite{Randall:1998uk,Giudice:1998xp} scenarios, etc.

In scenarios with a multiplicity of sectors which independently break supersymmetry, multiplicity of goldstini will be predicted~\cite{hep-ph:1002.1967}. Effects of supergravity  will induce a universal tree-level mass for the goldstini which is exactly twice the gravitino mass. Since the interaction strength of goldstini to visible sector fields can be greatly enhanced in comparison with  the (goldstino component of) gravitino, it may cause various new collider or cosmological effects~\cite{hep-ph:1004.4637}. We find that light goldstini dark matter from heavier goldstini decaying can account for the reported XENON1T anomaly.

This paper is organized as follows. In Sec ~\ref{sec:II}, we discuss the goldstini DM explanation of the XENON1T excess. In Sec ~\ref{sec:III}, numerical results of the fit to excess are given. Sec ~\ref{sec:conclusions} contains our conclusions.

\section{\label{sec:II}Very Long-lived Goldstini Dark Matter}
In GMSB, the LSP gravitino can act as the DM candidate. The light gravitino mass is predicted in GMSB by~\cite{hep-ph:9801271}
\beqa
m_{\tl{G}}\simeq \f{F}{k\sqrt{3}M_{Pl}}=\f{1}{k}\(\f{\sqrt{F}}{100{\rm TeV}}\)^2 2.4 {\rm eV}~,
\eeqa
with the model-dependent coefficient $k< 1$, and possibly $k\ll1$. Unlike the gravitino in gravity mediation, in which the interactions of gravitino will be of gravitational strength, the dominant gravitino interactions in GMSB come from its spin-1/2 component goldstino, whose derivative couplings are suppressed by $1/F$ and typically more important than the gravitational couplings suppressed by powers of $1/M_P$. The goldstino is the Goldstone fermion from spontaneous SUSY breaking. Its coupling with other fields is determined by the derivative coupling of goldstino to the supercurrent, much like the derivative coupling of pions in spontaneous chiral symmetry breaking\footnote{It is convenient to use the goldstino interactions in non-derivative form, which can be obtained in a given linearly realized SUSY model. The derivative and non-derivative forms of goldstino coupling are expected to give identical scattering amplitudes with a single external goldstino because the derivative coupling is part of the nonlinearly realized SUSY effective lagrangian, which can be obtained from the corresponding linearly realized SUSY model by field redefinition~\cite{hep-ph:9805512}.}.

 With multiple sector SUSY breaking, SUSY in each sector will be spontaneously broken at a typical scale $F_i$, yielding a corresponding goldstino.  One linear combination of goldstini will be eaten by the gravitino via the super-Higgs mechanism, the remaining goldstini fields are still propagating degree of freedoms. In the case where one or more SUSY
breaking sectors have direct couplings to the SSM to mediate SUSY breaking, the MSSM fields will actually couple more strongly to the goldstini than to the gravitino.
The interactions of the goldstini to MSSM chiral superfield $(\phi,\psi, F_\phi)$ are given by~\cite{hep-ph:1002.1967}
 \beqa
  {\cal L}_{\rm int}
  \supset \frac{1}{F_{eff}} \sum_{i,a}
    \frac{\tilde{m}_i^2 V_{ia}}{r_i} \zeta_a \psi \phi^\dagger -\frac{i}{\sqrt{2}F_{eff}} \sum_{i,a}
    \frac{\tilde{m}^\pr_i V_{ia}}{r_i}
    \zeta_a \sigma^{\mu\nu} \lambda F_{\mu\nu}.
\eeqa
where $F_i \equiv r_i F_{eff}$ ($\sum_i r_i^2 = 1$) and $\tl{m}_i^2$ ($\tilde{m}^\pr_i$) the contribution to soft scalar(gaugino) masses by each sector, respectively. One can see clearly the couplings of the goldstini in the case where there are only two SUSY breaking sectors with $F_1\gg F_2$.  The couplings to the uneaten goldstini is generically a factor of $F_{1}/F_2$ stronger than those to the gravitino for $\tl{m}_1^2\lesssim \tl{m}_2^2$ ($\tilde{m}^\pr_1\lesssim\tilde{m}^\pr_2$). The couplings of chiral superfields to goldstini can be approximately obtained from that of gravitino couplings by replacing $F_{eff}=\sqrt{F_1^2+F_2^2}\simeq F_1$ with $F_2$. Similarly, for hierarchical $F_i$, we can obtain approximately the low energy(below all $\tilde{m}_{i}^2$ ) local interaction term involving two matter fermions and two goldstini from that of the goldstino/gravitino~\cite{hep-th:9709111}
\beqa
{\cal L}_{eff}\supset-\f{1}{F_2^2}(\tl{\zeta}\sigma^\mu\pa^\nu\overline{\tl{\zeta}})(\bar{f}\overline{\sigma}_\nu\pa_\mu f)+\f{\al}{4F_2^2}(\tl{\zeta}\sigma^\mu\pa^\nu\bar{f})(\overline{\tl{\zeta}}\overline{\sigma}_\nu\pa_\mu f)~.
\eeqa
after integrating out the heavy fields involving sfermions etc.
Here $\al$ is a free parameter that can reproduce the linear realization results with $\al=-4$.
We can calculate the low energy $e^-\tl{\zeta}\ra e^-\tl{\zeta}$ scattering amplitudes to be
\beqa
&&\sum\limits_{spin}\left|{\cal M}\right|^2_{e^-\tl{\zeta}\ra e^-\tl{\zeta}}\\&=&\f{1}{F_2^4}\left\{ \[(m_e^2+m_{\tl{\zeta}}^2-s)+m_{\tl{\zeta}} m_e\]^2s^2+\[(m_e^2+m_{\tl{\zeta}}^2-u)+m_{\tl{\zeta}} m_e\]^2u^2-2m_{\tl{\zeta}}^2(t-2m_e^2)s u\right\}~,\nn
\eeqa
with the Mandelstam variables $s,t,u$ in the non-relativistic limit
\beqa
t&\approx& -2m_e T~,\nn\\
s&\approx& m_e^2+m_{\tl{\zeta}}^2+2m_e\(m_{\tl{\zeta}}+\f{1}{2}m_{\tl{\zeta}} v^2\)~,\nn\\
u&=& (m_{\tl{\zeta}}-m_e)^2-2m_e(\f{1}{2}m_{\tl{\zeta}} v^2-T ),
\eeqa
and  the special choice $\al=-4$. The amplitudes for $e^-\tl{\zeta}\ra e^-\tl{G}$ and $e^-\tl{G}\ra e^-\tl{G}$ will take similar forms and  be suppressed by additional $F_2/F_{eff}$ and $(F_2/F_{eff})^2$ factors in contrast to $e^-\tl{\zeta}\ra e^-\tl{\zeta}$.

It is interesting to note that the goldstini, although have masses twice of the LSP gravitino at tree level, can act as the DM candidate if the life-time of its decaying into gravitino is longer than the age of the universe. The lifetime of the goldstini $\tl{\zeta}$, which can later decay into gravitino $\tl{G}$ and neutrino pairs, can be estimated to be~\cite{hep-ph:1002.1967}
\beqa
\tau^{-1}_{\tl{\zeta}\ra \tl{G}\nu \bar{\nu}}\simeq \f{1}{128\pi^3}\f{m_{\tl{\zeta}}^9}{F_{eff}^4}\(\f{F_1}{F_2}\f{\tl{m}_2^2}{\tl{m}_1^2+\tl{m}_2^2}\)^2~
\simeq \[10^{22}~{\rm sec}
    \left( \frac{\sqrt{F_2}}{100~{\rm TeV}} \right)^4
    \left( \frac{100~{\rm GeV}}{m_{\tl{\zeta}}} \right)^7\]^{-1}.
\eeqa
So it is easy to adjust the various SUSY breaking parameters so as that the lifetime of $\tl{\zeta}$ is larger than $13.6 {\rm Gyr}$ and acts as a DM candidate. The $\tl{\zeta}$ DM can possibly scatter off the XENON electrons to generate the observed events in XENON1T experiment.

We can define the non-relativistic DM-electron interaction cross section at momentum transfer $q^2=\al^2_e m^2_e\approx 2 m_e T$  and DM form factor $F(q)$~\cite{hep-ph:1108.5383}
\beqa
\overline{\sigma}_e&=&\f{\mu_{\tl{\zeta} e}^2}{16\pi m_{\tl{\zeta}}^2 m_e^2}\left|{\cal M}(q)\right|^2_{q^2\approx (\al m_e)^2}~, \nn\\
\left|{\cal M}(q)\right|^2&=&\left|{\cal M}(q)\right|^2_{q^2\approx (\al m_e)^2} |F(q)|^2~,
\eeqa
with $\mu_{\tl{\zeta} e}$ the reduced mass for $e-\tl{\zeta}$ system. So, the velocity averaged differential ionization cross section for electrons are given by~\cite{astro-ph:1206.2644}
\beqa
\f{d}{d T} \langle\sigma v\rangle=\f{\overline{\sigma}_e m_e}{4\mu_{\tl{\zeta} e}^2}\int dv \f{f(v)}{v}\theta(v-v_{min})
\int_{q_-}^{q_+} \f{q dq}{\al^2m_e^2}  K(q,T) |F(q)_{DM}|^2~,
\eeqa
with
\beqa
q_\pm=m_{\tl{\zeta}} v\pm \sqrt{m^2_{\tl{\zeta}} v^2-2 m_{\tl{\zeta}} T }.
\eeqa
Here $K(q,T)$ is the atomic excitation factor and we take $K\simeq 0.1$ for electron recoil energy $T\sim 2$ keV. We assume a standard Maxwell-Boltzmann velocity distribution with a peak velocity of $v\simeq 0.1 c$ in $f(v)$.

The tree level masses of goldstini are twice of LSP gravitino, the thermal produced gravitino DM abundance will overclose our universe for $m_{3/2}>{\rm  keV}$. In fact, if LSP gravitinos are in thermal equilibrium at early times and freeze out at the temperature $T_f$, its relic density will be given by
\beqa
\Omega_{3/2}h^2=\f{m_{\tl{G}}}{\rm keV}\[\f{100}{g_*(T_f)}\]~.
\eeqa  We assume that some means of gravitino dilution will take place so as that the  gravitino dark matter will give negligible contributions to $\Omega_{DM} h^2$.

 To explain the XENON1T anomaly via DM, the DM needed to be boosted. There are many possibilities to boost the goldstini DM. We note that the most natural approach is to introduce multiple goldstini scenario(here we adopt two goldstini) so as that the long-lived Goldstini DM can be generated from the decay of other goldstini. We assume that the goldstini DM $\tl{\zeta}$ will be generated by the decaying of certain mother DM particle, for example, the other very long-lived goldstini  $\tl{\zeta}^\pr$ from multiple sector SUSY breaking scenarios which is slightly heavier than $\tl{\zeta}$. A tiny mass difference is already enough to boost $\tl{\zeta}$, for example, to $v=0.1 c$. So most of the current DM component is the long-lived goldstini (and its mother decaying DM particle $\tl{\zeta}^\pr$).

Both goldstini have degenerate tree-level masses $2m_{3/2}$. Their masses can be split slightly by different R-symmetry violating operators generated via higher loops.  We can choose proper mass mixing matrix $V_{ia}$ so as that the heavier goldstini $\tl{\zeta}^\pr$ has the $\tl{\zeta}^\pr-\psi-\tl{\phi}$ couplings approximately proportional to $m_{\tl{\phi}}^2/F_3$ while the lighter goldstini $\tl{\zeta}$ has the $\tl{\zeta}-\psi-\tl{\phi}$ couplings proportional to $m_{\tl{\phi}}^2/F_2$. The effective $F_{eff}$, which determines the goldstini and gravitino masses, can be much heavier $F_{eff}=\sqrt{F_1^2+F_2^2+F_3^2}\gg F_3\gg F_2$. Both goldstini can decay into LSP gravitino via $\tl{\zeta}_i\ra \tl{G}\bar{\psi}\psi$. However, the lighter goldstini  $\tl{\zeta}$ can still be stable by choosing proper $F_{eff}$. Then the heavier goldstini can decay dominantly into the lighter one because $F_{eff}\gg F_3$.
The decay width of $\tl{\zeta}^\pr\ra\tl{\zeta} \bar{\psi}{\psi}$ can be estimated to be
\beqa
\Ga_{\tl{\zeta}^\pr\ra\tl{\zeta} \bar{\psi}{\psi}}&\approx&\f{1}{128\pi^3}\f{m_{\tl{\zeta}^\pr}^9}{F_2^2F_3^2}~.
\eeqa
So the lifetime of $\tl{\zeta}^\pr$ can be estimated to be
\beqa
\tau_{\tl{\zeta}^\pr\ra\tl{\zeta} \bar{\psi}{\psi}}\approx 1.337\tm 10^{-15}\cdot\(\f{\sqrt{F_2}}{1 GeV}\)^4\(\f{\sqrt{F_3}}{1 GeV}\)^4\(\f{0.5 MeV}{m_\zeta}\)^{9} ~s\approx 4.0\tm 10^{17} s~.
\eeqa
So in order for $\tau_{\tl{\zeta}^\pr}$ to be $13.6$ Gyr, we should have
\beqa
\sqrt{F_2 F_3}\simeq (10~{\rm TeV})^2~.
\eeqa

The flux at the earth can be estimated to be~\cite{Buch:2020mrg}
\beqa
\f{d \Phi}{d E}=\f{d N_{\tl{G}}}{d E}\f{f_{\tl{\zeta}^\pr}}{4\pi \tau_{\tl{\zeta}^\pr} m_{\tl{\zeta}^\pr}} J_{\rm decay}~,
\eeqa
with $f_{\tl{\zeta}^\pr}$ the fraction of dark matter being $\tl{\zeta}^\pr$, which can be chosen to be ${\cal O}(1)$ here. The dominant contribution to the J-factor
\beqa
J_{\rm decay}=\int \rho_{\rm DM}(s)ds d\Omega~,
\eeqa
comes from the Milky Way halo, which is about $10^{23} {\rm GeV}/{\rm cm}^{2}$. So the maximal flux of goldstini DM $\tl{\zeta}$ on Earth from $\tl{\zeta}^\pr$ decaying can be estimated to be~\cite{Buch:2020mrg}
\beqa
\Phi_{\tl{\zeta}} \approx 4 \times 10^6 \, {\rm{cm}}^{-2}{\rm{s}}^{-1} f_{\tl{\zeta}^\pr} \left( \frac{4~MeV}{m_{\tl{\zeta}^\pr}} \right) \left( \frac{4 \times 10^{17} \, {\rm{s}}}{\tau_{\tl{\zeta}^\pr}}\right).
\eeqa
As the mass of  $\tl{\zeta}^\pr$ is nearly degenerate with(slightly heavier than)$\tl{\zeta}$, we adopt $m_{\tl{\zeta}^\pr}=m_{\tl{\zeta}}$ for simply in the flux estimation.

 The differential event rate for the DM scattering with electrons in xenon is given by
 \beqa
 \f{dR}{d T}=n_T \f{\Phi_{\tl{\zeta}}}{v}\f{d\langle\sigma v\rangle}{d T}.
 \eeqa

  The accompanied total neutrino and anti-neutrino flux is given by $\Phi_{\nu,\bar{\nu}}\simeq 2\Phi_{\tl{\zeta}}$. For simply, we assume that equal amounts of $\nu_e,\nu_{\mu},\nu_{\tau}$ are produced from the decaying of $\tl{\zeta}^\pr$ with $\Phi_{\nu_i,\bar{\nu}_i}=\f{1}{3} \Phi_{\nu,\bar{\nu}}$ and the average kinetic energies for neutrinos are identical to that of the light goldstini $E_{\tl{\zeta}}\simeq m_{\tl{\zeta}} v^2/2$, which is of order $2{\rm keV}$. Light goldstini and neutrinos from heavier goldstini decaying will stream freely after being produced from $\tl{\zeta}^\pr$ decay.

The neutrino-electron scattering cross section at low energy is given by~\cite{neutrino}
\begin{eqnarray}
\frac{d \sigma^{SM}_{\nu e}}{dE_R} &=& \frac{G_F^2 m_e}{2 \pi}\biggl[ (g_v + g_a)^2 +(g_v-g_a)^2\left(1-\frac{E_R}{E_\nu}\right)^2 + (g_a^2 - g_v^2)\frac{m_eE_R}{E_\nu^2} \biggr], \nonumber
\end{eqnarray}
where $G_F$ is the Fermi constant, $m_e$ is the electron mass, $E_R$ is the electron recoil energy and $E_\nu$ is the incoming neutrino energy. The $g_v$ and $g_a$ couplings depend on the neutrino flavor. For electron neutrinos we have
\begin{equation}\label{eq:gv_ga_e}
g_{v}^e = 2\sin^2 \theta_W + \frac{1}{2}; \, \, \, g_{a}^e  = +\frac{1}{2},
\end{equation}
while for muon and tau neutrinos
\begin{equation}\label{eq:gv_ga_mu}
g_{v}^{\mu,\tau} = 2\sin^2 \theta_W - \frac{1}{2}; \, \, \, g_{a}^{\mu,\tau}  = -\frac{1}{2},
\end{equation}
where $\sin^2\theta_W=0.23$ is the weak mixing angle. We neglect possible contributions from tiny neutrino magnetic momentum because we do not refer to a enhanced magnetic momentum explanations of the XENON1T excess in this paper. As expected, numerical results indicate that neutrino fluxes from goldstini decay products will give negligibly small contributions without enhanced magnetic momentum.

\section{\label{sec:III}Numerical Results}

In order to compare our results with the data reported by the XENON1T collaboration, the resulting differential events rate should be smeared by a Gaussian distribution
\beqa
\f{d R}{d T_{m}}=\int d T \f{d R}{d T}\f{1}{\sqrt{2\pi}\sigma}e^{-\f{(T-T_m)^2}{2\sigma^2}}\al(T),
\eeqa
with variance~\cite{Aprile:2020yad}
\beqa
\sigma=T_m \(\f{31.71}{\sqrt{T_m}}+0.15\)~,
\eeqa
and the total efficiency $\al(T)$ presented in FIG.2 of ~\cite{Aprile:2020tmw}. Here $T_m$ is the measured recoil energy.

 Then we build a likelihood,
\begin{equation}
\mathcal{L}(s_i) = \prod_{i=1}^{29} \frac{(b_i+s_i)^{d_i} e^{-(b_i+s_i)}}{d_i!}
\label{eq:like}
\end{equation}
where $s_i$ and $b_i$ are the binned signal and $B_0$ the background predictions, $d_i$ is the observed counts. The background predictions and observed counts are took from FIG.~4 of ~\cite{Aprile:2020tmw}, with an exposure of 0.6473 tonne-years~\cite{Athron:2020maw}. For the background only hypothesis, we obtain $\chi^2_b=-2\ln\mathcal{L}(s_i=0)=43.9$.

\begin{figure}
\centering
\includegraphics[width=.49\textwidth]{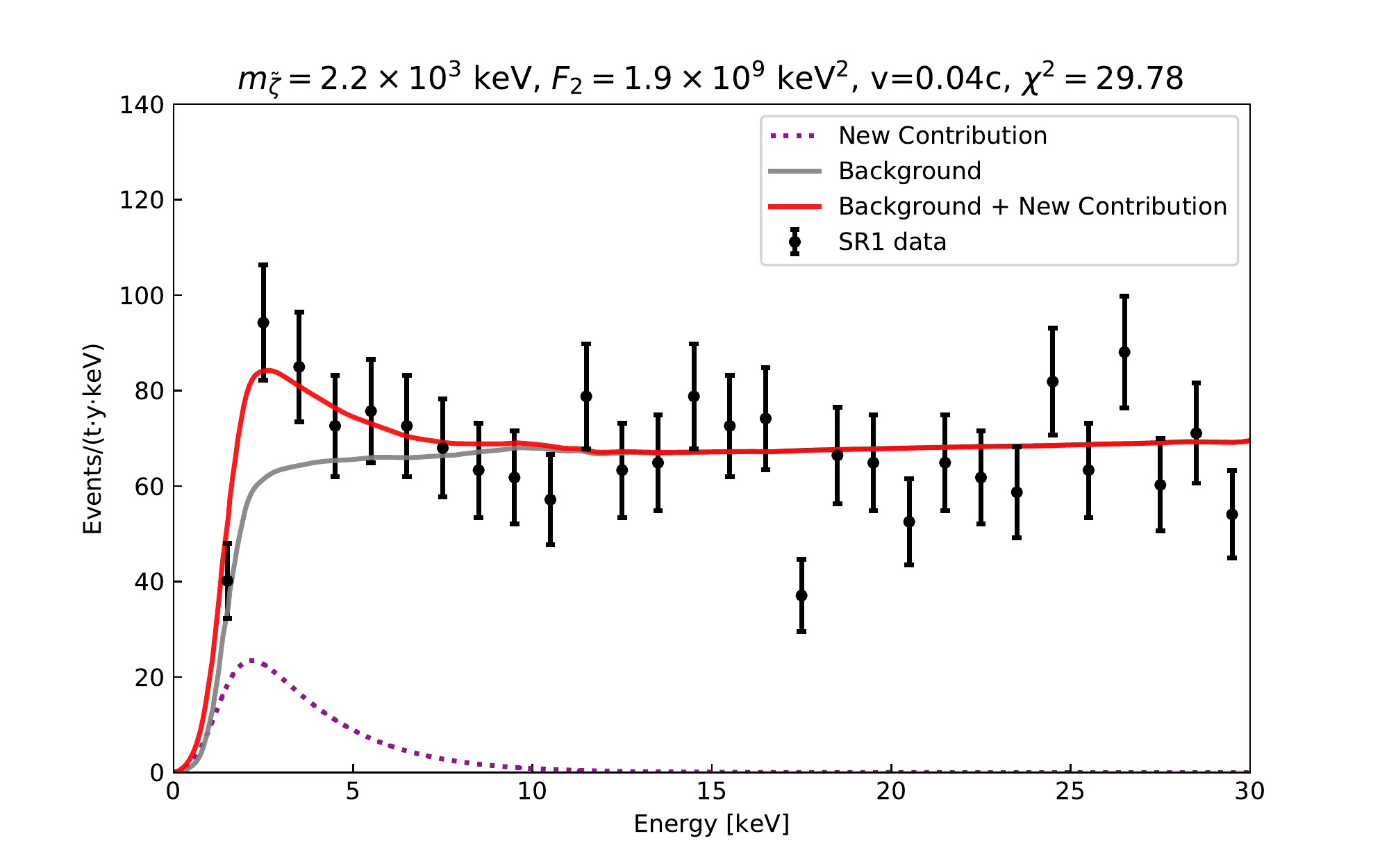}
\includegraphics[width=.49\textwidth]{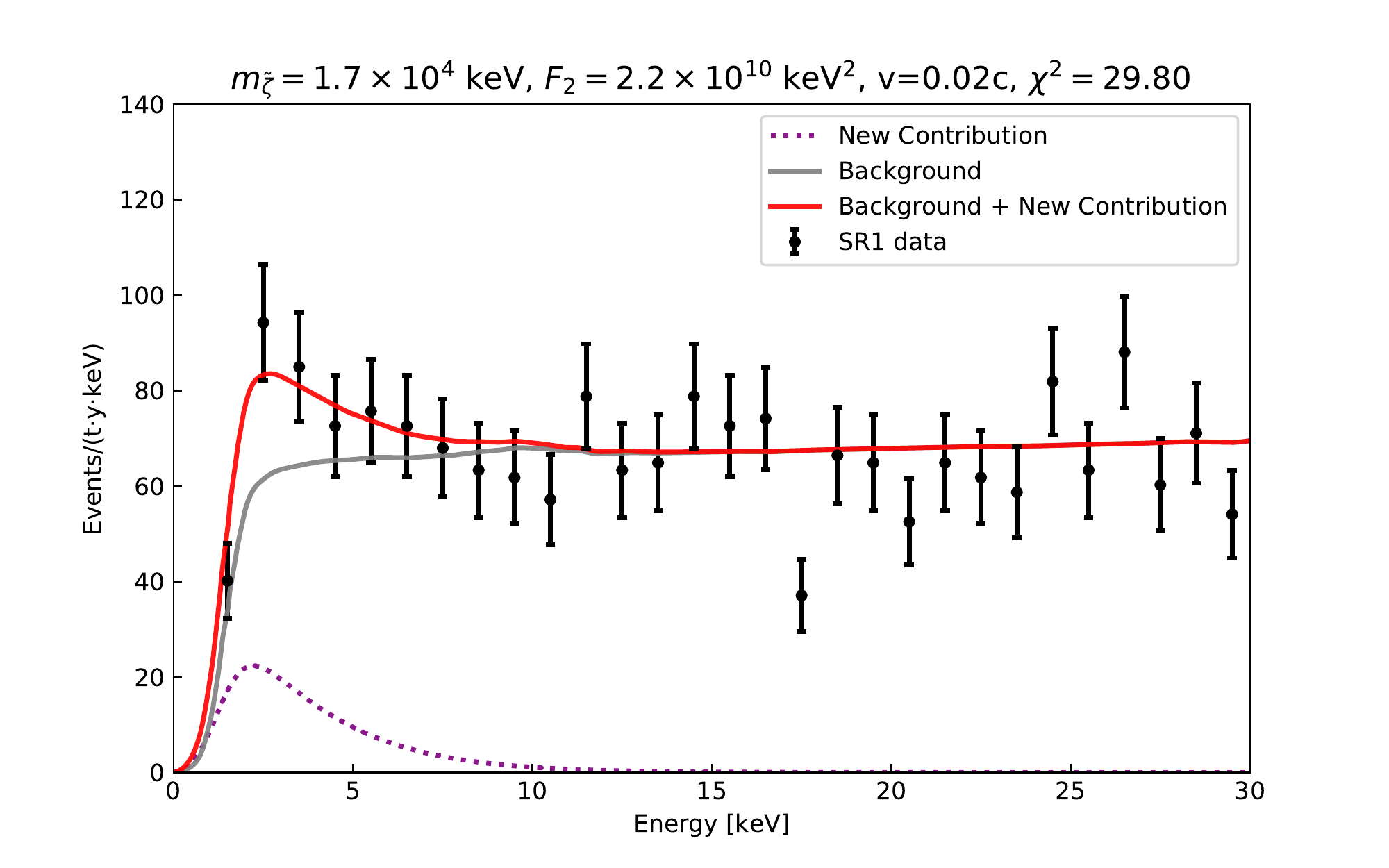}
\caption{The low-energy electron recoil spectrum of signal (purple dash-line) and signal+background (red line) for our best-fit point, compared to background model (gray line) and observed data(black dot with error bar). The best-fit point is $F_2=1.9\times 10^{9}~{\rm keV}^2$, $m_{\tilde{\zeta}}=2.2\tm 10^{3}~{\rm keV}$ and $v=0.04~c$ with $\chi^2_{\rm best}=29.78$(the left panel) and $F_2=2.2\times 10^{10}~{\rm keV}^2$, $m_{\tilde{\zeta}}=1.7\tm 10^4~{\rm keV}$ and $v=0.02~c$ with $\chi^2_{\rm best}=29.80$(the right panel), respectively. }
\label{fig:best_fit}
\end{figure}

We perform a random scan in the parameter space
\begin{equation}
F_2 \in [10^8, 10^{11}]~{\rm keV}^2, ~~~  m_{\tilde{\zeta}}\in[50, 10^5]~{\rm keV}, ~~~ v \in [0.001,0.3]~c.
\label{eq:scan}
\end{equation}
We show the two best-fit points in FIG.\ref{fig:best_fit}, which are favored over the background only hypothesis by $\Delta \chi^2 = \chi^2_b-\chi^2_{\rm best} = 14.116$ and $14.102$, respectively.
In FIG.\ref{fig:best_fit}, the recoil energy spectra of the background model and best-fit signal predictions are shown, as well as the observed counts. We can see that our interpretation can fit the XENON1T data fairly well.

\begin{figure}
\centering
\includegraphics[width=.80\textwidth]{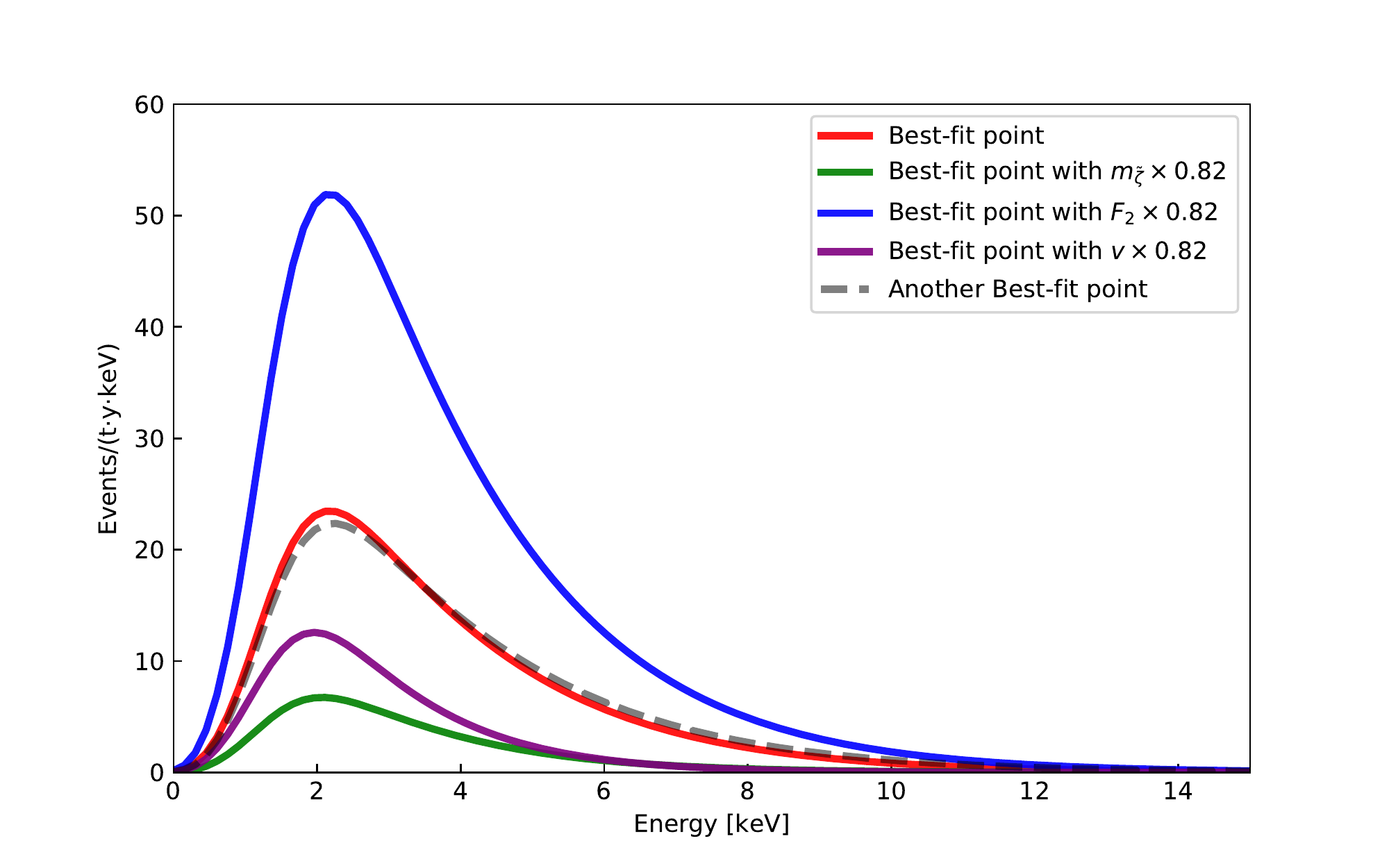}
\caption{The dependence of signal prediction on the variation of parameters for best-fit point $F_2=1.9\times 10^{9}~{\rm keV}^2$, $m_{\tilde{\zeta}}=2.2\tm 10^{3}~{\rm keV}$ and $v=0.04~c$ (red line). The other colourful curve corresponds to the variation of one parameter with other parameters fixed. "Another best-fit point" (gray dash line) in this figure refers to $F_2=2.2\times 10^{10}~{\rm keV}^2$, $m_{\tilde{\zeta}}=1.7\tm 10^4~{\rm keV}$ and $v=0.02~c$.}
\label{fig:var}
\end{figure}

The dependence of signals on the input parameters are shown in Fig.\ref{fig:var}.
Each curve of Fig.\ref{fig:var} corresponds to the variation of one parameter with other parameters fixed. We can see from the figure that 
increasing the velocity and the mass of DM particle will increase the incident kinetic energy while decreasing the value of $F_2$ will increase the interaction strength, all of which can increasing the signals. Meanwhile, tuning them simultaneously can obtain similar electron recoil spectrum to the best-fit point.

\begin{figure}
\centering
\includegraphics[width=.49\textwidth]{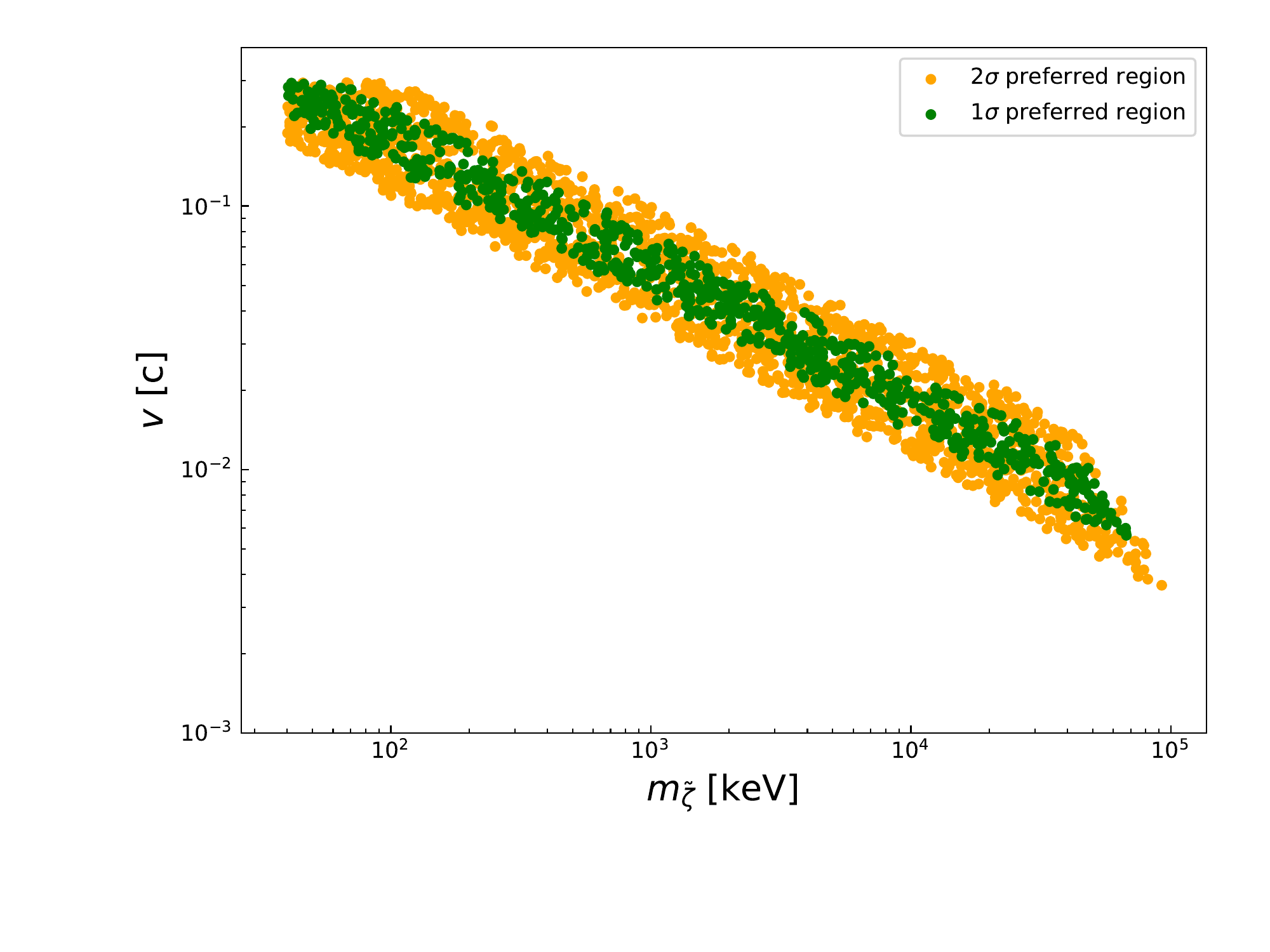}
\includegraphics[width=.49\textwidth]{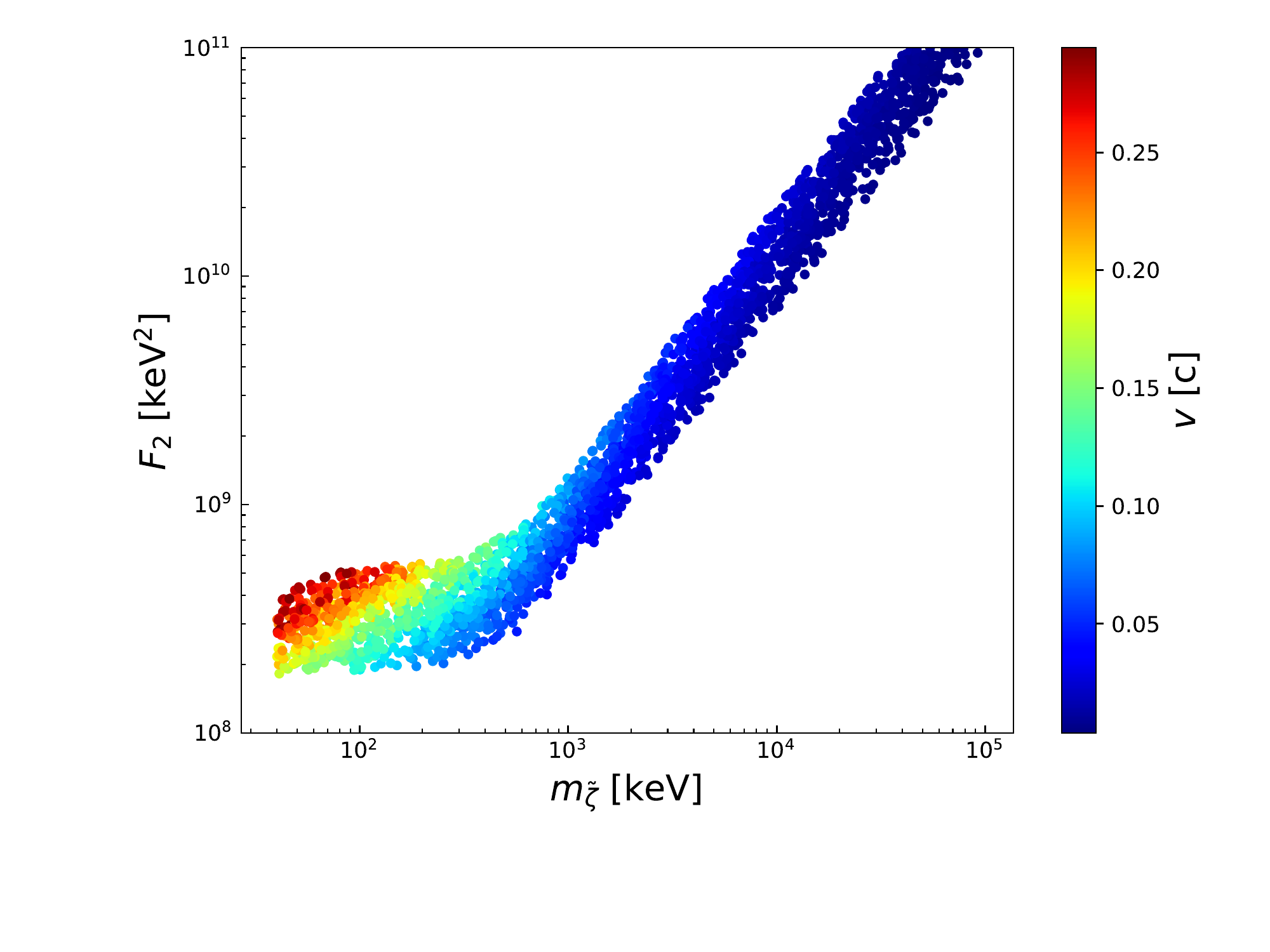}
\caption{The preferred parameters on $m_{\tilde{\zeta}}$ versus $v$ plane (left) and $m_{\tilde{\zeta}}$ versus $F_2$ plane (right). In the left panel, the points preferred within $1\sigma$ and $2\sigma$ confidence level are colored  green and orange, respectively. The points within $2\sigma$ confidence level  are  shown in the right panel with colors indicating the velocity $v$. }
\label{fig:scan}
\end{figure}

Furthermore, we present the preferred parameter regions in FIG.\ref{fig:scan}. To explain the XENON1T excess, the DM velocity $v$ should decrease with increasing $m_{\tilde{\zeta}}$, as depicted in the left panel of FIG.\ref{fig:scan}. Larger value of $m_{\tilde{\zeta}}$ in general needs smaller DM velocity. The smallest DM velocity needed is about $0.003c$, which is still a bit larger than the escape velocity of DM $0.0015 < v_{esc} < 0.002$ from the Milky Way. So, the boost from heavier goldstini decay is necessary to interpret the excess.

The right panel of FIG.\ref{fig:scan} shows that $F_2$ should increases with increasing $m_{\tilde{\zeta}}$. As a result, we find $F_2>1.0 \tm 10^8 \kev^2$ for $m_{\tilde{\zeta}}>40 \kev$. It is also interesting to note that, in order to explain the XENON1T excess with fixed $m_{\tilde{\zeta}}$, increasing the DM speed needs to increasing the value of $F_2$, which corresponds to a decreased interaction strength.
\section{\label{sec:conclusions}Conclusion}
 In the scenario with a multiplicity of sectors which independently break supersymmetry, multiplicity of goldstini are predicted. We propose a new interpretation of the electron recoil excess at 2-7 keV observed in the XENON1T experiment with very long-lived goldstini DM elastically scattering off the electrons. The goldstini DM can be boosted by the late-decay of the other nearly degenerate (long-lived) goldstini DM, with their tiny mass difference being converted into kinetic energy of the lighter goldstini DM and neutrinos. It is also possible for inelastic scattering of goldstini DM off an electron, in which the goldstini can convert into gravitino DM. This process and the gravitino elastic scattering process are both highly suppressed in compare with the elastic scattering process of light goldstini.  We show that viable parameter space can be found which can explain the excess of electron recoil events around 2-3 keV recently reported by the XENON1T experiment.

We should note that for light goldstini of order several keV, it is also possible to explain the excess by the transition magnetic momentum of goldstini-gravitino. We will discuss such a possibility in our subsequent studies.

\begin{acknowledgments}
 This work was supported by the Natural Science Foundation of China under grant numbers 11575053, 11675147 and 11775012, by the Australian Research Council through the ARC Centre of Excellence for Particle Physics at the Tera-scale CE110001104.
\end{acknowledgments}

\end{document}